\documentclass[11pt]{article}
\usepackage{color,amsfonts,amsmath,amssymb,mathrsfs,verbatim,graphicx}

\def\esi{\mbox{E}_6}
\def\ese{\mbox{E}_7}
\def\ee{\mbox{E}_8}
\def\soot{\mbox{SO}^+(1,3)}

\def\rrr{{\mathbb{R}}}
\def\bv{\mbox{\boldmath $v$}}

\def\suth{\mbox{SU}(3)}
\def\uo{\mbox{U}(1)}

\def\fots{\mbox{\raisebox{+1pt}{\footnotesize{$\frac{1}{3}$}}}}
\def\ftts{\mbox{\raisebox{+1pt}{\footnotesize{$\frac{2}{3}$}}}}

\def\vgaps{-10pt}
\def\vgapb{-7pt}
\def\vgapa{-4pt}
\def\vgapq{-5pt}
\def\vgapc{-3pt}
\def\vgapt{-2pt}
\def\vgapp{-1pt}

\def\setb{\setlength{\baselineskip}{0.625\baselineskip}}

\begin{document} 

{\setlength{\baselineskip}{0.625\baselineskip}

\begin{center}

      {\LARGE {\bf  Unification through Generalised Proper Time: \\ 
                      \vspace{4pt} 
           The Short Story }}

   \vspace{13pt}

 
 \mbox {{\Large David J. Jackson}} \\ 
  \vspace{5pt}  
  {david.jackson.th@gmail.com}  \\

  
 \vspace{12pt}
 
 { \large October 28, 2020 }

 \vspace{40pt}

{\bf  Abstract}

\vspace{-10pt} 
 
\end{center}


     The central arguments for a generalised form of proper time as the basis
    for a unified theory, accounting for the Standard Model of particle physics and encompassing all four forces of nature including a consistent `quantum gravity', are presented. After first outlining a set of criteria to be ideally met by any proposal for a comprehensive unification scheme we describe the extent to which
     a theory based on generalised proper time satisfies these objectives. In comparison with other approaches we also discuss the change in perspective regarding the relations between space, time and matter associated with this theory.

\vspace{40pt}

{






}



\section{The Ideal of Unification}
\label{stor1}
\vspace{\vgapt}

\begin{quotation}
    It appears therefore that certain phenomena in electricity and magnetism lead to the same conclusion as those of optics, namely, that there is an {\ae}thereal medium pervading all bodies,\ldots
\vspace{\vgapq}
\flushright{James Clerk Maxwell (\cite{Maxem} part I.15, 1865)} 
\end{quotation} 
\vspace{\vgapp}    

    Through the history of physics advances towards a more unified worldview have been generally considered particularly  significant markers of progress. This is the case for Newton's law of universal gravitation, the innovations of Faraday and Maxwell culminating in the theory of electromagnetism and the Glashow-Weinberg-Salam  model of electroweak unification. 
    Such landmark developments have inevitably led to the ongoing quest for a unified theory of all forces of nature and all fundamental physics, from the smallest scale studied in the laboratory to the largest scale observed in cosmology (see for example~\cite{Pen} final chapter, \cite{Hooft}). 
     
     Opinion naturally varies concerning whether such a unified theory might actually exist, and if it does whether it might actually be attainable, but regardless of that standpoint the nature of what unification should entail is
      relatively uncontroversial. Hence while the ambition and means of constructing such a theory might be contentious the \textit{ideal} of unification itself is less ambiguous, as outlined in the following five criteria.
     Within minor variations in individual points these  
       are intended as an objective account  of the features desired in a unified theory.
        
\begin{itemize}

\item[1.] {The basis of \textit{uni}fication should be \textit{one simple} idea or entity universally pervading all physical phenomena and from which all such phenomena can be derived, at least in practice at an elementary level. In particular, the theory should not be constructed or designed to directly model empirical observations.}
\vspace{\vgapc}

\item[2.] {The basic idea or entity should be \textit{fundamental}, in the sense of an elementary and  intuitively reasonable source and bedrock for unification. That is, the basic entity should not be too unfamiliar, contrived or invented in a manner that would stand in need of explanation in itself as a manifest `loose end' of the theory.}
\vspace{\vgapc}

\item[3.] {The basis of unification should have a direct \textit{mathematical} expression leading to the unique development of a rigorous and well-defined physical theory.  The theory should provide an account of why \textit{this} mathematics should describe the \textit{physical} world.}
\vspace{\vgapc}

\item[4.] {The theory thus constructed should be capable of accounting for existing successful theories and models in the appropriate limits and of explaining a non-trivial range of the extensive existing \textit{empirical} data concerning the elementary structure of matter. This should include a consistent union of general relativity with quantum theory and an account of the Standard Model of particle physics and of the observations underlying the standard cosmological model.}
\vspace{\vgapc}

\item[5.] {As the theory develops there should be no intractable contradictions with the physical world and  connections should be made with all aspects of fundamental phenomena without any gaps in explanatory power. Further, beyond what is currently empirically known the theory should make testable \textit{predictions}.}   
\vspace{\vgapc}

\end{itemize}     

     In the following section we introduce the idea of employing \textit{time} as the basis for unification and describe the extent to which the development of a theory founded upon generalised proper time meets the above desired criteria for a unified theory. 
     In the final section, again with reference to the above criteria 1--5, we consider how the need for a significant change in perspective might hinder the adoption of the new theory despite the successes.


\vspace{\vgaps}

\vspace{-1pt}

\section{Generalised Proper Time as a Unifying Basis}
\label{stor2}
\vspace{\vgapt}

\begin{quotation}
    {\bf Space} and {\bf time} are commonly regarded as the {\bf forms} of existence of the real world, {\bf matter} as its {\bf substance}.
\vspace{\vgapq}    
\flushright{Hermann Weyl (\cite{Weyl2} opening of Introduction, 1922)} 
\end{quotation}    
\vspace{\vgapp}

    The three central entities for any physical theory are \textit{space}, \textit{time} and \textit{matter}. Most of physics is based upon posited forms of \textit{matter} in a background arena of space and time. Such postulated matter acts as a substratum underlying physical phenomena as might be theoretically derived for comparison with empirical observations.
      In general these theories are essentially \textit{phenomenology} in adopting a form of matter explicitly to \textit{model} the observations. This is largely the case for the Standard Model of particle physics, with matter in spacetime introduced in the form of the various fields of the theory as related through a specific Lagrangian~\cite{Teub}.
      However, while almost all science places the focus on matter a unified theory, in seeking a \textit{deeper explanation} of observed phenomena, might be expected to be of a different nature. Otherwise the question can ever be raised concerning a possible further supporting substratum of matter at the next level down or regarding the ultimate origin of `matter' itself.
   
     Of the above three basic entities one such approach to unification places the initial emphasis on \textit{space} rather than on \textit{matter}, and in particular on extra spatial dimensions over and above the familiar \mbox{4-dimensional} spacetime arena (see for example~\cite{Liu} and references therein).
      With structures of matter deriving from the properties of the higher-dimensional spacetime, together with the assumptions needed to extract physics in \mbox{4-dimensional} spacetime, this offers a more unifying framework albeit still with a somewhat complex starting point.
       Further, despite the initially encouraging proposal of Kaluza and Klein for unifying gravity with electromagnetism in the 1920s (\cite{Liu} section~II.A), modern theories with extra dimensions of space  have not achieved any decisive success in accounting for the Standard Model (see for example~\cite{Witt, Jitt}) or other empirical phenomena and it remains the case that `there is no direct evidence that there are extra spatial dimensions'  (\cite{Liu} section~V).
     
      Given that of the three basic entities forms of \textit{matter} and forms of \textit{space} have been extensively researched as a potential primary basis for a unified theory, it is perhaps not unreasonable to consider the third basic entity, and hence `forms of \textit{time}', as a basis for unification -- that is, if the construction of such a theory is  \textit{possible}. 
      In light of the five key criteria sought for unification highlighted in the previous section
      we describe how this is indeed the case for the proposal  of a generalised form of proper time as the basis of a comprehensive unified theory in the following five subsections.
      
\vspace{\vgapb}     
      
\subsection{Simplicity}
\label{stor21}

\vspace{\vgapa}

   The aim to `keep it simple' urges against basing a unified theory upon any \mbox{non-trivial} construction, such as specific matter fields in a  4-dimensional spacetime background
    whether postulated directly or deriving from a specific higher-dimensional spacetime structure. All such frameworks incorporate \textit{time} as can be identified with the choice of a one-dimensional timelike  progression parametrising the evolution of all physical phenomena. The structure of time \textit{itself} as a one-dimensional continuum \textit{alone} then clearly offers a much \textit{simpler} starting point for a theory.
   
     Unlike the endless varieties of possible postulated forms of matter and a vast range of possible higher-dimensional spacetime geometries, the simple one-dimensional progression in time provides a far more unique and unambiguous basis for a unification scheme, 
      employing minimal assumptions about the nature of the physical world.
 The objective of deriving both \textit{matter and space} from a simple basis in \textit{time} alone evidently presents a more \textit{unifying} proposal.
     
 \vspace{\vgapb}      
  
\subsection{A Fundamental Basis}
\label{stor22}

\vspace{\vgapa}     

    It is through the concept of \textit{motion} that the three basic entities of space, time and matter are intimately connected (\cite{Weyl2} Introduction). 
   Laws of physics, in particular those that have proved successful, are typically formulated in terms of dynamical equations for the evolution of matter with respect to a temporal parameter, with the spacetime arena infused throughout by a flow of time. While the equations of motion, the forms of matter and the structure of spacetime will vary, from this objective point of view time is then \textit{fundamental} to any description of events in the world, and in accompanying all change it is perhaps the most  fundamental concept in all of science.
   
    Time is also arguably the most fundamental element of our subjective encounters with the physical world -- indeed we cannot even \textit{think} about physics without a progression in time to connect our thoughts. 
     These objective and subjective aspects of time are combined in all experiments and observations we can perform in the empirical world.
       As an innate and irreducible feature of such observations time hence provides a very \textit{conservative} starting point for a unified theory, as something we are intimately familiar with unlike novel forms of matter or extra dimensions of space. 
     The question then naturally concerns how forms of matter in space might conceivably \textit{derive} from a fundamental basis in time alone.
  
 \vspace{\vgapb}      
  
\subsection{Mathematical Expression}
\label{stor23}

\vspace{\vgapa}     
        
   Given the very simple nature of time as the fundamental entity for a unified theory considered here the question of identifying \textit{any}  sufficiently rich mathematical structure with the potential to describe the physical world is immediately evident.
      We begin with the one-dimensional continuum of progression in time which is structurally isomorphic to the ordered continuum of real numbers $\rrr$, which does give an intrinsic and direct connection with an elementary mathematical structure. This allows the introduction of  a real time parameter:
\begin{equation}
  \label{sinr}
              s \, \in \, \rrr
\end{equation}

    As alluded to in the previous subsection the 4-dimensional spacetime manifold of general relativity, with local inertial coordinates $(x^0,x^1,x^2,x^3)$ constructed about any point, can be considered to be infused with a flow of time as might be measured by a local `clock'. A local infinitesimal interval of proper time
     $\delta s \in \rrr$, invariant under local Lorentz transformations, can then be expressed as:
\begin{equation}
  \label{sxxxx}
   (\delta s)^2 \, = \, (\delta x^0)^2 -  (\delta x^1)^2 
                          -  (\delta x^2)^2 -  (\delta x^3)^2
\end{equation}  
  
   The key observation for the present theory, in \textit{beginning} with time $s \in \rrr$ in equation~\ref{sinr}, is that equation~\ref{sxxxx} can be \textit{interpreted} as a possible basic arithmetic expression \textit{for} an infinitesimal interval  $\delta s \in \rrr$, consistent with the elementary axiomatic properties of the real line. Under this change in perspective, hinging upon the reading of equation~\ref{sxxxx}, rather than time flowing \textit{through} the pre-existing spacetime of the right-hand side the initial focus is instead on the left-hand side with 
      time intervals  
    identically expressed in this 4-dimensional quadratic form acting as the \textit{progenitor} of the local spacetime geometry itself.
    This deployment of the \textit{arithmetic} substructure implicit in the \mbox{one-dimensional} flow of time to represent the \textit{geometric} structure of
    4-dimensional spacetime  is illustrated by the progression from figure~\ref{sopen}(a) to figure~\ref{sopen}(b).   

  
   Incorporating the local $4\times 4$ metric $\eta = \mbox{diag}(+1,-1,-1,-1)$ of a Lorentzian spacetime the direct expression in equation~\ref{sxxxx} for a proper time interval, in the form 
    $(\delta s)^2  = \eta_{ab} \delta x^a \delta x^b$ for $a,b = 0,1,2,3$ and 
    with Lorentz symmetry $\soot$,
  can be further arithmetically generalised to the $n$-dimensional 
   homogeneous $p^{\mathrm{th}}$-order  form
   (again summing over repeated indices, now for   $a,b,c,\ldots = 0,\ldots,n-1$):   
\begin{equation}
 \label{salpha}
  (\delta s)^p  \, = \, \alpha_{abc\ldots}\delta x^a 
                            \delta x^b \delta x^c \ldots
\end{equation}
   Here  in general $p>2$, each coefficient $\alpha_{abc\ldots} = -1,0$ or $+1$ and
     a larger group $\hat{G}$,  generalising from the Lorentz group,  acts as symmetry transformations upon the set of $\{\delta x^a\}$ components leaving $\delta s$ invariant.
    
  As a generalisation from equation~\ref{sxxxx}, incorporating a 4-dimensional quadratic Lorentzian form as a substructure, equation~\ref{salpha} involves additional components over and above those needed to describe the local 4-dimensional spacetime geometry. 
   We noted above that equation~\ref{sxxxx} can be interpreted as time not just flowing through a local spacetime but as \textit{generating} the spacetime geometry itself as described for figure~\ref{sopen}(b). Similarly the extra components of the form of proper time in equation~\ref{salpha}, residual to the construction of 4-dimensional spacetime, can be interpreted as not just associated with or infusing matter in spacetime but as actually \textit{giving rise to} the structure of matter \textit{itself}, as we have attempted to represent in figure~\ref{sopen}(c).
     
\begin{figure}[htbp]  
\centering
\leavevmode
\includegraphics[width=14.4cm]{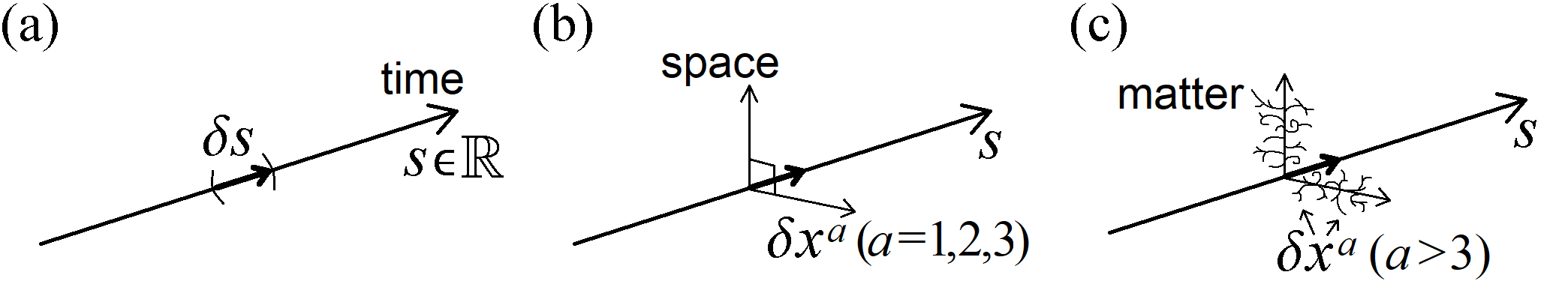}
\caption{\setb (a) The continuum of time of equation~\protect\ref{sinr} incorporates infinitesimal intervals
   $\delta s  \in \rrr$  (exaggerated in the diagram) that can be directly arithmetically expressed in
   \mbox{(b) the  form} of equation~\protect\ref{sxxxx} interpreted  as the \textit{origin} of local 4-dimensional spacetime with three spatial components and (c) the further generalisation to the form of equation~\protect\ref{salpha} with the excess components \textit{generating} structures of matter in spacetime.}
\label{sopen}
\end{figure}      
        
   Given the direct connection from the conceptual basis of the theory in the continuum of time as identified with the real line in equation~\ref{sinr} through the intrinsic arithmetic generalisation  via equation~\ref{sxxxx} to equation~\ref{salpha} as the basis for the structure of 4-dimensional spacetime with a matter content, the relation between the \textit{mathematical structure} of the theory and the \textit{physical structure} of the empirical world rests on a firm and substantial basis~\cite{Struct}. 
    With the flow of time permeating all physical phenomena this theory hence 
   exhibits a thorough and secure
    grasp on the structure of matter.
  
 \vspace{\vgapb}      
  
\subsection{Physical Realisation}
\label{stor24}

\vspace{\vgapa}     

The  additional components of equation~\ref{salpha} over 4-dimensional spacetime, as represented by the short curved lines in figure~\ref{sopen}(c), can be interpreted as a basis for `matter fields' in an extended spacetime, in principle with the potential to  reproduce or account for existing successful field theories such as the Standard Model of particle physics.     
  The question then concerns how such a framework  actually performs when measured against empirical facts about the world.
   In developing a full physical theory from generalised proper time it is convenient to rewrite the general expression for infinitesimal intervals in equation~\ref{salpha} in terms of the components
 $v^a := \frac{\delta x^a}{\delta s}   {\big{\vert}}_{\mbox {\tiny $\delta s \! \to \! 0$}}$ 
   of an $n$-vector $\bv_n$
   in the form:
\begin{equation}
  \label{lpvn}
  L_p(\bv_n)_{\hat{G}} 
  \; := \; \alpha_{abc\ldots} \frac{\delta x^a}{\delta s}
   \frac{\delta x^b}{\delta s}\frac{\delta x^c}{\delta s}\ldots
   \Big\vert_{\delta s \to 0} \; = \;
    \alpha_{abc\ldots}v^a v^b v^c \ldots \; = \; 1
\end{equation}
  This notation includes the homogeneous power $p$, dimension $n$ and full symmetry
   $\hat{G}$ of this generalised form for proper time  (\cite{TimeE} equation~43, \cite{Gener} equation~13).
   
    On fixing $p=2$  extensions beyond the case \mbox{$(p=2,n=4,\hat{G}=\soot )$}
   of $(n\! =\!4 )$-dimensional spacetime    through 
      $n=5,6,7,\ldots$ with a  $\hat{G} = \mbox{SO}^+(1,n\! -\! 1)$
      symmetry are uniformly permitted, as might be associated with extra spatial dimensions. However given that we do not \textit{perceive} any spatial properties for the additional components the restriction to a quadratic $p=2$ structure for the general case is an unnecessary assumption. For $p>2$ particular values of $n$ and specific groups $\hat{G}$, subsuming 
      the 4-dimensional Lorentzian form and symmetry, are mathematically favoured in describing a high degree of symmetry   (\cite{TimeE} equations~48--49, 52--54 and 63--64, \cite{Gener} equations~15--23, 30--31 and 34 and associated references). This inevitably leads to analysis of equation~\ref{lpvn}
    for a \mbox{$(p=3,n=27,\hat{G}=\esi )$} case  and in turn for a 
      \mbox{$(p=4,n=56,\hat{G}=\ese )$} case, hence via 
     $\esi$ and $\ese$  incorporating a central role for the unique series of exceptional Lie groups
      and also the octonion algebra~\cite{Octo}.
     
     The symmetry $\hat{G}$ of the full $n$-dimensional form of proper time in equations~\ref{salpha} and \ref{lpvn} is broken through the necessary extraction of the local 4-dimensional spacetime substructure on the right-hand side of equation~\ref{sxxxx}. This leaves the surviving symmetry:
\begin{equation}
 \mbox{Lorentz} \times G \, \subset \, \hat{G}
 \label{gbreak}
\end{equation} 
    as a product of the local external Lorentz symmetry $\soot$ and a local internal gauge symmetry $G$ as the basis for physics on an extended 4-dimensional spacetime manifold
      (\cite{TimeE} subsection~4.1,  \cite{Gener} equations~28--29, \cite{KKone} subsection~2.3).
    
     The symmetry breaking for the $\hat{G} = \ese$ level leads to a corresponding fragmentation of the components of $\bv_{56}$ for this $n\!=\!56$ case of equation~\ref{lpvn} generating matter fields transforming under the broken symmetry on the left-hand side of equation~\ref{gbreak}. The properties directly identified for these matter field fragments include Dirac spinor structures under the external
      Lorentz symmetry, singlets and triplets under an internal $\suth_c \subset G$ identified as a colour gauge symmetry, a $1$:$\ftts$:$\fots$ fractional charge structure under an internal $\uo_Q \subset G$ as associated with electromagnetism and an intrinsic left-right asymmetry as well as further elements of electroweak symmetry breaking (\cite{TimeE} section~4 figure 4, \cite{Gener} subsection~3.1  equation~35--36).
      
      These features hence closely reproduce a large part of the properties of a generation of leptons and quarks in the Standard Model.  
   In contrast with the case for a foundation in forms of matter or forms of a higher-dimensional spacetime the means of building a theory based on proper time alone are much more restrictive.
    Given this very simple basis the observation that it is even \textit{possible} to construct a theory starting from the concept of time alone, as outlined for figure~\ref{sopen} 
    and equations~\ref{lpvn} and~\ref{gbreak} is of some significance in itself. That the essentially unique and minimal manner in which a physical theory of matter in  4-dimensional spacetime can be derived  from an  analysis of the symmetry breaking of
    generalised proper time
       actually \textit{works} in leading so \textit{directly} to highly non-trivial connections with the esoteric empirical structures of particle physics as briefly reviewed above then seems particularly striking.
    
    These Standard Model-like structures are identified in the local symmetry breaking over a 4-dimensional spacetime substructure as described for equation~\ref{gbreak}. Since we are beginning with time alone the nature of the matter composition more generally in constructing an extended external spacetime manifold consistent with the constraints implied in the general form of proper time of equation~\ref{lpvn} can also be investigated. In this manner structures can be identified coherently combining both the principles of  general relativity and the phenomena of quantum theory, hence 
     incorporating a candidate for a consistent  scheme of `quantum gravity'~\cite{QGrav}.
     
       The theory can be interpreted as a generalisation from general relativity
       to a unifying framework encompassing an essentially classical theory of gravity
       and from which a standard formalism of quantum theory might be recovered as a limiting case
        (\cite{Gener} subsection~5.1, \cite{QGrav} subsection~4.1). 
        The characteristic quantum property of indeterminism arises from an intrinsic local degeneracy in the matter field composition in solutions for a continuous extended 4-dimensional spacetime geometry,
      with an element of non-locality inherent in constructing an extended space itself.  
          An account can be given of the origin of discrete particle quanta  from the constraints of the theory 
         (\cite{QGrav} subsections~4.2 and 6.1).       
     Long-standing conceptual issues such as the nature of the interface between quantum and classical systems and the `measurement problem' can also be
     addressed
      (\cite{QGrav}  figures~1, 2, 4 and 5, subsections~2.1, 7.3 and 7.4).     
            
\vspace{\vgapb}         
  
\subsection{Completeness and Predictions}
\label{stor25}

\vspace{\vgapa}     

 As alluded to in the abstract and criterion~4 of section~\ref{stor1}, the main targets of a unified theory involve a single coherent origin for the four forces of nature (the relation of gravity to internal gauge forces is described in~\cite{KKone}), an explanation of the Standard Model of particle physics (see for example~\cite{TimeE}), a consistent amalgamation of gravitation with
     quantum theory~\cite{QGrav} and  an account of the standard cosmological model (in particular regarding the dark sector and the very early universe~\cite{Unifi} chapters~12~and~13). The above references describe the progress that has been made in these areas for this theory based on generalised proper time and the directions for further work.

 At this stage in the development of the theory  there are no evident contradictions either internally or when compared with our current knowledge of the empirical world.
  While the explanatory power is still advancing
there are also no evident areas of fundamental physics which are of concern as seemingly ultimately beyond the reach of this unified theory. However  further development is still needed  to fully reproduce quantum and particle phenomena in the laboratory.
 Further progress and understanding will also be needed
  to address theoretical issues 
  such as the singularities of black holes and the Big Bang, together with the thermodynamic properties of the former and the nature of the early universe evolving from the latter, requiring the full expression of a quantum gravity framework (\cite{QGrav} subsections~2.3 and 7.1). 
 In all cases the theory is at least capable of framing an approach to address the associated questions.

        In particular, in light of the need to account for a full three generations of leptons and quarks, with a corresponding full spinor structure and incorporating a full electroweak theory, the progress described in the previous subsection points to the proposed construction of a form for proper time in equation~\ref{lpvn} for the case \mbox{$(p=8, n=248, \hat{G}=\ee )$}.
        The employment of  a symmetry action of uniquely the largest exceptional 
        Lie group $\ee$ involving the algebra of octonion-valued elements~\cite{Octo} in this currently hypothetical full form for proper time to potentially account for the full Standard Model in the symmetry breaking over a \mbox{4-dimensional} spacetime manifold is a \mbox{\textit{mathematical prediction}} of the theory as described in detail in~\cite{TimeE}.
          Constraints on the possible form this mathematical structure could take in turn lead to provisional \textit{physical predictions} for new phenomena in the areas of neutrino, \mbox{Higgs and dark} matter physics~(\cite{Gener} section~4). With the focus of the theory on explaining and going beyond known laboratory phenomena such predictions could be within reach of empirical verification within the foreseeable future and hence provide tests for the theory.


\vspace{\vgaps}

\section{A New Perspective for Unification}
\label{stor3}
\vspace{\vgapt}

\begin{quotation}
    Perhaps what we mainly need is some subtle change in perspective -- something that we all have
     missed \ldots.
\vspace{\vgapq}
\flushright{Roger Penrose (\cite{Pen} final chapter closing, 2004)} 
\end{quotation}
\vspace{\vgapp}

    While the concept of matter as an independent substance in space and time, employed to organise and comprehend our observations of the physical world, has been of great pragmatic value in science it has fallen short as the basis for a comprehensive unified theory. 
    Similarly the change in emphasis to unification schemes based on extra dimensions of space has resulted in limited empirical success.
      For the present theory both matter and the space through which it is perceived to flow are constructed through substructures of the multi-dimensional form for proper time of equation~\ref{lpvn} as the underlying unifying principle as introduced in the discussion of figure~\ref{sopen}.   
      With reference to the opening quote of section~\ref{stor1} time
       is a very familiar entity that also 
       exhibits the property of  `pervading all bodies' in that all phenomena necessarily \textit{happen in time}.
      This framework has a simple and fundamental basis while also being able to reach a broad range of elementary empirical phenomena as required of a unified theory. The theory hence scores significantly higher regarding the ideal of unification compared with matter-based theories or models based on extra spatial dimensions as measured across criteria 1--5 outlined in section~\ref{stor1}, as  reviewed in subsections~\ref{stor21}--\ref{stor25}. Hence,  on providing a novel opportunity to explore unification based on time as the fundamental entity, the question might be raised concerning why such an approach has not already been widely investigated. 
      
       Given the progress made for unification schemes based on matter or extra spatial dimensions, with an  
   extensive range of conceivable hypothetical structures and significant technical challenges encountered, 
    a  pressing need has been to build upon the existing developments to improve the empirical connections and hence with a focus upon criteria 3--5 of section~\ref{stor1}.    
         This has perhaps led to criteria 1 and 2 being afforded somewhat less attention as a basis in either `matter' or `space' has been presupposed or adopted at a much earlier stage of these developments. 
           However in pursuing a better match with the empirical world for \mbox{criterion 4}  there is perhaps a danger that once the input structure of a theory becomes too complex, beginning for example with a more contrived link with the Standard Model, then a divergence from the ideal of simplicity and  
   criteria 1 and 2 may open up. That is, rather than the `one-to-many' explanatory power sought for unification a theory may in the limiting case exhibit more of a `many-to-many' match between its input basis and the empirical world, becoming more a case of phenomenology rather than a unified theory.
   Theories with extra spatial dimensions, while \textit{plausible} in starting with an additional structure over 4-dimensional spacetime to account for matter, are perhaps
    typically of this nature~(\cite{Witt,Jitt}, see also discussion in \cite{Gener} subsection~2.1). 
  
    Compared to the shift in worldview from matter to space the further transition to time as the fundamental entity perhaps marks an even more significant change in perspective on the elementary structure of the world. 
    However while the notion of time is very simple and familiar, very much consistent with the ideal of unification described for criteria~1 and 2, it is highly counter-intuitive that a rich and comprehensive unified theory with broad explanatory power \textit{could} derive from such an elementary basis.
   That this seeming initial \textit{implausibility} may  act as  an obstacle to the adoption of this approach 
     is in some sense perhaps paradoxical in that it is essentially an archetypical feature \textit{sought} for a unification framework.
      That is, the fact that generalised proper time scores highly on criteria~1 and 2 ironically may make  the theory readily overlooked as being \textit{too} ideal for unification (see also \cite{Gener} subsection~5.2).
    
      As the one-dimensional continuum of equation~\ref{sinr} incorporating the intrinsic arithmetic substructure  of equation~\ref{salpha} time possesses the \textit{dual}  properties of simplicity and compositeness as specifically desired in a unification scheme with a potential `one-to-many' explanatory power. 
      While at an elementary local level models with \mbox{extra spatial dimensions} place the focus on the right-hand side of equation~\ref{sxxxx} and add further quadratic terms, here the change in perspective to placing the focus upon time on the left-hand side essentially amounts to the rather subtle observation
       that  this expression for proper time 
        can be generalised to that of equation~\ref{salpha} with the \textit{assumption} of a quadratic structure with $p=2$ dropped, as noted after equation~\ref{lpvn}.
     This more general case with  $p>2$ leads directly to a richer local mathematical structure which is found to exhibit close links with the Standard Model 
      as  reviewed in subsection~\ref{stor24}, \textit{without} needing to subsequently contrive these properties.
    
 Similarly as for deriving \textit{matter} from \textit{space}, here the  central conceptual idea that \textit{matter} and the \textit{space} in which it is located \textit{can} derive from \textit{time} alone
  hinges on the interpretation of mathematical expressions. 
   It is not possible to 
   `see' how this could work by starting with any \textit{visualisation} since space itself is a derived entity. However our immediate encounters with the physical world, and the familiar manner in which we might instinctively think of it, are with forms of matter already distributed in space and evolving in time.
     Hence the initial continuous flow of time and corresponding properties of the  real line of equation~\ref{sinr} as represented in figure~\ref{sopen}(a)
      might be misleadingly interpreted as already flowing through a pre-existing space,  as could not be `pictured' in any other way. 
    That would miss a main point of the theory and render redundant the local \textit{construction}  of space \textit{from} time, 
    via a reading of equation~\ref{sxxxx} from  left to right 
   as   \textit{then} pictured in the \textit{resulting} \mbox{3-dimensional} space as intended to be indicated in  figure~\ref{sopen}(b).
    On adopting the new perspective
      the further generalisation of a proper time interval to the intrinsic arithmetic substructure in equations~\ref{salpha} and \ref{lpvn} then leads to the identification of a matter content in spacetime via the symmetry breaking of equation~\ref{gbreak} as initially described for  figure~\ref{sopen}(c)
  (and elaborated in \cite{Struct}).  
    
  Given the uniqueness of the starting point in equation~\ref{sinr}, and also of the explicit mathematical structures for equation~\ref{lpvn} with symmetries described by exceptional Lie groups discussed in subsection~\ref{stor24}, the question might be raised concerning \textit{why} the intermediate form on the right-hand side of equation~\ref{sxxxx} should also be highlighted as underlying the geometric frame through which we observe the world.
    It is this 4-dimensional Lorentzian form that precipitates the symmetry breaking and
     leads directly to a series of empirical connections as also briefly summarised in  subsection~\ref{stor24}. The suggestion is: firstly the prominence of a \textit{quadratic} form is \textit{needed} to perceive any world at all in a \textit{space} with locally Euclidean properties, and secondly for the \mbox{3-dimensional} spatial case of  equation~\ref{sxxxx} the resulting physics is rich enough to support observers in the form of ourselves. This is considered a mild anthropic argument, of the level of finding ourselves for example living on Earth rather than on Venus. That is, of a limited number of possibilities for constructing a universe we naturally populate a habitable case (for further discussion of uniqueness see~\cite{Unifi} section~13.3).
   
    A readiness or otherwise to adopt this new perspective with a foundation in time, rather than in forms of matter or extra spatial dimensions for example,  with the great simplification of initial assumptions implied, to some extent will depend upon the level of conviction in criteria 1 and 2 for the ideal 
     of unification outlined in section~\ref{stor1}. In any case the unification scheme described in this paper requires a significant  reassessment of our conceptions of space, time and matter and the relations between them.
    Compared with other unification schemes on adopting the new perspective this theory also performs very well for criteria 3--5 in making highly non-trivial inroads in directly accounting for the Standard Model as well as providing a framework for quantum gravity.
     While naturally further work is needed to further develop and understand the theory (building upon~\cite{TimeE,Gener,QGrav} for example), the successes achieved in meeting significant elements of all five desired criteria for unification stand in favour of establishing generalised proper time as a suitable basis for a unified theory.


\vspace{\vgaps}

{\small
{\setlength{\baselineskip}{0.62\baselineskip}

\par}
}


\par}

\end{document}